\documentstyle[prl,aps,epsf]{revtex}
\twocolumn
\tighten

\begin{document}
\draft
\twocolumn[
\hsize\textwidth\columnwidth\hsize\csname@twocolumnfalse\endcsname
\title{Magnetoresistance induced by low-field control of phase
separation in La$_{0.5}$Ca$_{0.5}$MnO$_{3}$}
\author{F. Parisi$^{a,+}$, P. Levy$^{a,++}$, L. Ghivelder$^{b}$,
G. Polla$^{a}$, D. Vega$^{a,+}$}
\address{
(a) Departamento de F\'{\i}sica, Comisi\'{o}n Nacional de Energ\'{\i}a At\'{o}mica,
Avda. Gral Paz 1499 (1650) San Mart\'{\i}n, Buenos Aires, Argentina  \\
(b) Instituto de F\'{\i}sica, Universidade Federal do Rio de
Janeiro, C.P. 68528, Rio de Janeiro, RJ 21945-970, Brazil
}

\maketitle

\begin{abstract}
The effect of low magnetic fields on the transport properties of a
manganite compound with magnetic phase separation is studied. The
different behavior obtained in samples of
La$_{0.5}$Ca$_{0.5}$MnO$_{3}$ related to the way in which the low
field is applied is consistent with a picture of changes in the
metallic fraction induced by the magnetic field in a field-cooled
cycle.  Using a simple model of conduction through a binary mixture,
the interrelation between magnetoresistance, metallic fraction and
percolation temperature is accounted for. A new physical coefficient
relating magnetic field and metallic fraction emerges as the relevant
parameter in the description of phase separated manganites.\\
\end{abstract} 
]
\narrowtext

\section{Introduction}

The hole-doped rare-earth manganites L$_{1-x}$A$_{x}$MnO$_{3}$, where
L is a lanthanide and A a divalent alkaline earth, display a wide
variety of spin, charge, and orbital states.\cite{ref1} The close interplay
between them gives rise to different electronic, magnetic and
structural phases, such as paramagnetic insulator (PI), ferromagnetic
metallic (FMM) or charge ordered antiferromagnetic insulator (COAF)
phases. \cite{ref1},\cite{ref0}. That interrelation leads to many unusual physical properties, of which the negative magnetoresistance (MR) effect is perhaps the
most spectacular one. \\
The MR effect has been extensively studied, \cite{ref1} and it has been associated, in general, to a magnetic transition. In particular, the MR of a
homogeneous and single crystalline system displaying a PI to FMM
transition shows a peak close to the transition temperature T$_{C}$.
This was explained on the basis of the double-exchange mechanism, as
the result of the thermally induced spin disorder suppression by an
external magnetic field, with the consequent shift in T$_{C}$. Far
below T$_{C}$ the magnetic field has no influence in the
(homogeneous) state of the system, and the MR is negligible below
T/T$_{C}<$ 1/2. For granular manganites, instead, the effect of the
magnetic field is appreciable even at very low temperatures, due to
the role played by intergrain barriers.\cite{ref2} A similar
scenario accounts for homogeneous systems showing a FMM-COAF
transition. In this case, the presence of a low magnetic field can
only introduce a little shift in T$_{co}$, with a consequent peak in
MR; neither single crystals nor granular materials show an appreciable value of the MR far below T$_{co}$. However, at
fields high enough to induce the COAF-FMM transition (or to suppress
the FMM-COAF one) the MR can achieve high values in a wide
temperature range below T$_{co}$.\cite{ref3}\\ 
The above described
scenario has now a novel and unexpected ingredient: the fact that the
low temperature state of some of the manganites consists of
coexisting ferromagnetic (FM) and COAF phases has introduced new
degrees of freedom into the problem. In this context, the physical
properties of the system are strongly dependent on the fraction of
the FM phase and its spatial distribution (percolative or not) to
account for transport behavior. \\
Despite the fact that the reason for
the existence of the phase separated (PS) state is not yet well
understood, theoretical\cite{ref4} and
experimental \cite{ref5,ref6,ref7,ref8,ref9} results agree that the
competition between different phases, which is resolved in a short
length scale, is very sensitive to the particular kind of disorder in
the compound: site A disorder  in
La$_{5/8-y}$Pr$_y$Ca$_{3/8}$MnO$_3$,\cite{ref5} site B disorder in
Nd$_{0.5}$Ca$_{0.5}$Mn$_{0.98}$Cr$_{0.02}$O$_3$\cite{ref6} and 
Pr$_{0.5}$Ca$_{0.5}$Mn$_{1-x}$Cr$_{x}$O$_{3}$,
\cite{ref7} grain size effects in
La$_{0.5}$Ca$_{0.5}$MnO$_{3}$,\cite{ref8} or intra-granular strain
in Pr$_{0.7}$Ca$_{0.3}$MnO$_{3}$.\cite{ref9} 
The conduction through percolative
paths of the FM phase in PS systems has opened a new scenario for the
MR effect. The fourfold orders of magnitude in the drop of the
resistivity below 100 K found in La$_{5/8-y}$Pr$_y$Ca$_{3/8}$MnO$_3$
under a magnetic field as low as 0.4 T is far away from typical values
of the low field MR in homogeneous systems. The role played by the
magnetic field in this phenomena was interpreted as related to the
interplay between the percolative nature of the phase separation and
the domain alignment mechanism, since the field was considered too
low to modify the coexisting phase balance.\cite{ref5} \\
Crucial effects of the
magnetic field on PS systems suggesting a different interpretation
were recently reported. In La$_{0.7}$Ca$_{0.3}$MnO$_{3}$ scanning tunneling
microscopy experiments\cite{ref10} gave direct evidence of changes
of the FM fraction by the applied magnetic field for temperatures
just below T$_{C}$. In
Nd$_{0.5}$Ca$_{0.5}$Mn$_{0.98}$Cr$_{0.02}$O$_{3}$,\cite{ref6} it has
been shown that the low temperature fraction of the FM phase is
highly dependent on the magnetic field $H_{ann}$ under which the
system is cooled (annealing field): a
non-percolative-percolative transition is induced for H$_{ann}\ge $ 2
T, with the consequent changes of several orders of magnitude in the
resistivity. That transition can be only accounted for by the change
of the FM fraction $f$ induced by the magnetic field, a fact also
reflected in the linear $H_{ann}$ dependence of the low temperature
magnetization M.\\
The effects of applying moderate magnetic fields (H $\ge$ 5 T) has been also studied in other PS compounds. It has been shown that a magnetic field of 6 T produces the melting of the CO phase in Nd$_{0.5}$Sr$_{0.5}$MnO$_{3}$, inducing a structural transition and rendering the material FMM mostly. \cite{Rao} The possibility of controlling the relative phase fractions with magnetic fields was also claimed in Ref \cite{babush}. They have shown that different FM fractions can be tuned as a function of H in $^{18}$O rich samples of (La$_{0.25}$Pr$_{0.25}$)$_{0.7}$Ca$_{0.3}$MnO$_3$ when H exceeds a temperature dependent value H$_c$ of a few teslas. Specific heat measurements on charge ordered compounds Pr$_{1-x}$Ca$_{x}$MnO$_{3} $(0.3$\le x \le 0.5$) reveals also that a magnetic field H=8.5 T is able to modify the relative phase fractions.\cite{smolya}\\

The gap existing between the magnitude of the magnetic field used in the low field MR work (0.4 T), \cite{ref5}, and those needed for inducing changes in the relative fractions of the coexisting phases ($> 2$ T) \cite{ref6},\cite{Rao},\cite{babush} seems to enforce the idea of two different mechanisms to explain the MR effects in PS systems: domain alignment effect at low fields (under 1 tesla), the fraction change at moderate ones.

The aim of this paper is to give some insight into this subject by analyzing the simple idea that even when applying a low magnetic field in a field-cooled experiment it is possible to affect the fraction of the coexisting phases, a process we call field-induced fraction enlargement. We have performed MR measurements on La$_{0.5}$Ca$_{0.5}$MnO$_{3}$, a suitable compound for studying PS effects due to its PI-FM-COAF phase transitions sequence \cite{shiffer} and the coexistence of the COAF phase with nanodomains of the FM one below T$_{co} \approx$ 150 K (on cooling).\cite{mori}
As we previously reported, \cite{ref8} changing the grain size through
consecutive thermal treatments allow us to control the fraction of FM
and COAF phases. This fact
opens the possibility for studying the magnetic field effects for
different FM fractions below T$_{co}$. The measurements were performed with low magnetic fields, in a way that we claim the two effects, alignment (suppression of thermal disorder, intergranular
coupling, domain orientation) on one hand, and field-induced fraction enlargement on the other one, are clearly distinguished. Doing this, we could
establish the relevance of the fraction enlargement mechanism over
the magnetic domain alignment in percolative PS systems at low field. With this
result in hands, we analyze the general issue of MR in PS systems,
and show that the relation between MR and the FM fraction can be
accounted for within this framework with a simple model of conduction
by a binary mixture. \\ 

\section{Experimental}

Polycrystalline samples were obtained by a citrate/nitrate decomposition method as described in
 Ref. \cite{ref8}. Powder x-ray diffraction was used to check the phase purity; unit cell dimensions and structural parameters were analyzed using the Rietveld method. Four probe resistivity measurements were performed in the temperature range 30-300 K on polycrystalline pellets previously pressed and sintered. The two samples used in this work are those labeled C and EII in Ref. \cite{ref8}; the FM fractions were determined through magnetization measurements following the procedure explained there. 

\section{Results}

In Fig. 1 we show the resistivity ($\rho$)curves for the
two samples under study as a function of temperature.
Sample I, with 55$\%$ of FM phase at low temperatures, displays a
metallic behavior below 80 K, indicating the existence of percolative
paths of the FM phase. In sample II, with 9$\%$ of FM phase at low
temperatures and an insulating behavior, no signal of transport
percolation was found down to 40 K, were its resistance exceeded our
instrumental detection limit. 
In Fig. 2a we sketch two curves showing
the magnetoresistance MR = [$\rho(0)-\rho(H)]/\rho(H)$ of sample I,
with each result corresponding to a particular measurement process.
In the first procedure, hereafter referred to as field-cooled (FC)
mode, the low field (H = 0.65 T) was applied throughout the thermal
cycle as the sample is cooled.  In the second one, the field was
turned on and turned off periodically (TOTO mode) while cooling the
sample; the data points were obtained after the field was fully
established in each step. These procedures were used with the aim of
distinguishing the two above mentioned mechanisms involved in the
field dependence of the resistivity: the alignment and the
field-induced fraction enlargement effects. 

If the local free energy
describing the FM and COAF states is adequately accounted for by a
double-well potential with almost identical minimums, the main idea
is that a low magnetic field can, in the FC mode, prevent the
formation of otherwise COAF regions. In this case both mechanisms are
expected to be present in the FC mode, and this fact would be
reflected by the MR. On the other hand, if a low magnetic field is
applied after the fraction of the phases was established, the energy
barrier between the wells would prevent the change of the populations
if the field is applied for a short time. Therefore the only effect
in the resistivity would be related to domain orientation. Following
this argument, we have taken a period of 60 seconds for the switching
off the field in the TOTO mode, which is short compared with
relaxation times observed through similar processes in another PS
system,\cite{ref6} and also confirmed by careful relaxation experiments
performed on our samples. \\  
\begin{center}
\begin{figure}[tbp]
\hspace{-2cm}
\epsfxsize=4.50cm
\centerline{\epsffile{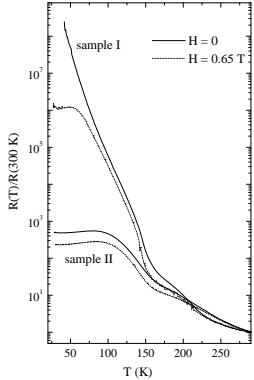}}
\caption{Resistivity at $H$=0 and 0.65 T (in the field
\label{Fig1} 
}
\end{figure}
\end{center}
\begin{figure}[tbp]
\epsfxsize=6.0cm\centerline{\epsffile{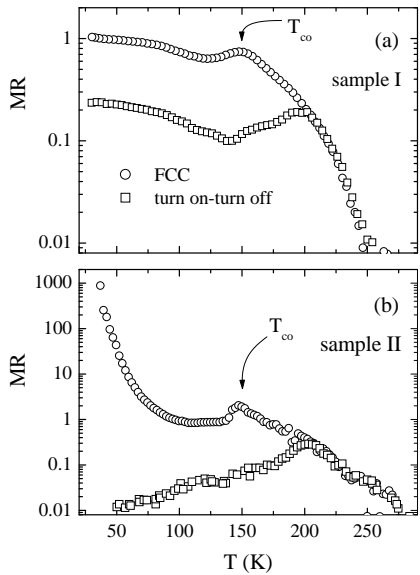}}
\caption{ a)Magnetoresistance [$\rho(0)-\rho(0.65 T)]/\rho(0.65 T)$ of the percolative PS sample I in the
field cooling mode and in the turn on - turn off field mode. b) The same
as in a) for the non-percolative PS sample II.
\label{Fig2}}
\end{figure}   
Comparing the temperature dependence of
the obtained MRs measured on cooling with the different procedures on
sample I, plotted in Fig. 2a, we can see that both MR have their
onset around T$_C$, and the curves increase close together with
decreasing temperature down to T$\approx$ 200 K. Bellow this temperature both
curves depart. Remarkably, this temperature is close to that at which
the onset of a structural phase separation has been reported.
\cite{ref12} The TOTO curve has a peak near T$_C$, and below
T$_{co}$ increases almost linearly towards a low temperature value of
0.25, resembling the  behavior of an homogeneous granular FMM
compound.\cite{ref13} On the other hand the FC curve rises steeply
below T$_C$, reaches a peak (MR=0.75) around T$_{co}$, and also
increases almost linearly towards a low temperature value as high as
1.2. \\
The different response of the compound to the magnetic field
regarding the way it was applied gives us a starting point for the
understanding of the MR effect in PS systems. When the role of the
magnetic field is only that of magnetic moments alignment, as in the
TOTO mode,  the MR of a PS compound is, irrespective of its PS
nature, the same as that of an homogeneous FMM one, due to the
intensive character of the MR.  The split of the curves when the
compound becomes unstable against phase separation and the large
values of the MR in the FC mode point to the fact that the low
magnetic field is able to modify the relative volumes of the
coexisting phases when applied throughout the thermal cycle. In this
situation the resistivities with and without applied field correspond
to media with different amount of FM phases, which is a distinctive
feature that can only  be obtained in PS systems. \\
In order to
support the  above described scenario we have measured the MR in the
FC and  TOTO modes on sample II (Fig. 2b). This sample is just below
the percolation threshold of the FM phase, as the lack of metallic
behavior at low temperatures indicates (see Fig.1). The MR in the
TOTO mode goes to zero for low temperatures, as expected in a
non-percolative sample.  However, the applied low field in the FC
mode induces an increase of the fraction of the FM phase, leading to
its percolation, with the consequent change of more than two orders
of magnitude in resistivity below 60 K. Thus, the low field plays an
unbalancing role in the FC mode, inducing a qualitative change, from
non-percolative to percolative transport, giving the colossal values
of the MR displayed in Fig. 2b. It is also worth noting that the
differences found in the MR figures obtained by the FC mode for
samples I and II seem to follow a definite relation among the FM
fractions and the MR values:  high MR are obtained  when the sample
has a low FM content. \\

\section{Discussion}

The presented results point to the fact that
field-induced enlargement of the FM phase is responsible for the
observed low-field MR in this PS perovskite. In the discussion that
follows we show that this fact is a suitable candidate to give an
explanation for the relation between MR and FM fraction at a given
temperature. We also show that within this framework it is possible
to account for the MR values at the insulator-metal transition
temperature (T$_{im}$) in the prototypical PS system
La$_{5/8-y}$Pr$_y$Ca$_{3/8}$MnO$_3$, an experimental observation not
yet properly understood.\cite{ref5} \\
To do this we use as main
framework the General Effective Medium theory,\cite{ref14} which
describes the resistivity $\rho_e$ of a binary mixture through the
relation
\begin{equation}
f\frac
{(\rho_e^{1/t}-\rho_{FM}^{1/t})}{(\rho_e^{1/t}+A_c\rho_{FM}^{1/t})}+
(1-f)\frac{(\rho_e^{1/t}-\rho_{co}^{1/t})}{(\rho_e^{1/t}+A_c\rho_{co}^{1/t})}=
0
\end{equation}
where $\rho_{FM}$ and $\rho_{co}$ are, respectively, the
resistivities of homogeneous FM and CO phases, explicitly including
their particular field and temperature dependence and granular
effects, $t$ is the critical exponent (we assume $t$ = 2), $f$ is the
fraction of the FM phase, and $A_c=1/f_c - 1$, where $f_c$ is the
critical percolative fraction of the  FM phase. \\
Through Eq. 1
we can obtain an expression for the MR of a PS system in terms of the
resistivities of the constitutive mediums and the corresponding FM
fraction $f$, under the assumption that the magnetic field induces
small changes in $f$.  As $\rho_{FM}$, $\rho_{co}$ and $f$ are, in
general, temperature and field dependent, we obtain an intricate
dependence of the MR with the external parameters. An important
simplification can be introduced by neglecting the dependence of
$\rho_{FM}$ with the magnetic field (the MR arising from that
dependence is, at most, 10$\%$ for low fields at T$_{co}$, as shown
by the TOTO curves of Fig. 2). With this approximation, the
calculated MR is defined by the ratio of the resistivities of the
sample when the FM fraction is changed by $\Delta f$, so that
\begin{equation}
MR=\frac {\rho_{e}(f, \rho_{co}/\rho_{FM})}{\rho_{e}(f+\Delta f,
\rho_{co}/\rho_{FM})}-1
\end{equation}
where $\Delta f$ is the actual field-induced fraction enlargement
value. It is worth noting that Eq. (2) is, in some sense, a
generalization of the MR concept, as it describes a MR obtained
between resistivity curves at fields $H$ and $H+\Delta H$, where in
this case the FM fraction enlargement effect is induced by $\Delta
H$. \\
We may now compare the predictions arising from this simplified
model with the experimental results of the MR of two prototypical
PS compounds: La$_{0.5}$Ca$_{0.5}$MnO$_{3}$ and
La$_{5/8-y}$Pr$_y$Ca$_{3/8}$MnO$_3$. As a distinctive feature, in the
case of La$_{0.5}$Ca$_{0.5}$MnO$_{3}$ we can study the MR in samples
with different values of $f$ at a constant temperature, for instance,
the temperature at which the MR peaks. In this case the value of
$\rho_{co}/\rho_{FM}$ is fixed, and the only unknown parameter is the
field dependence of $\Delta f$. In order to gain some insight into
the relation between $f$ and $H$ we have measured the resistivity of
samples I and II for several magnetic fields $H$ in the FC mode,
between 0 and 0.8 T, and calculated the MR at the peak temperature,
T$\approx$ 150 K, for all the intervals $\Delta H$ between these
fields. With this procedure we obtain the MR for different values of
$\Delta f$ as a function of $\Delta H$. The zero field data was ruled
out in order to partially exclude the neglected field effect on
$\rho_{FM}$. The results are shown in Fig. 3a. The calculated curves
were obtained  following the experimental results of Ref.
\cite{ref6}, assuming a linear relation between $f$ and $H$, i.e., $f
= f_0+\alpha_fH$, where $f_0$ is the zero field FM fraction and
$\alpha_f$ a field-independent factor, hereafter referred to as the
{\it fraction expansion coefficient}. We have taken the value
$\rho_{co}/\rho_{FM}$ =1000 and the zero field FM fraction of the
samples I and II from Ref. \cite{ref8}. The only free parameter for the
adjustment of the experimental results is $\alpha_f$. The $\alpha_f$
values obtained, 0.20 and 0.11 T$^{-1}$ for samples I and II
respectively, indicate a small dependence of $\Delta f$ with the
initial FM fraction $f_0$.  The agreement between the experimental
data and calculated curves reveals that the simplified model we are
dealing with is able to account for the relation between the MR and
FM fraction at a constant temperature, giving a quantitative basis
for the understanding of the MR effect in PS systems. \\
In order to
obtain  an additional insight on the results that are expected for PS
systems in the fraction enlargement scenario, we have used Eq. 2 to
describe the MR effect occurring in
La$_{5/8-y}$Pr$_y$Ca$_{3/8}$MnO$_3$, a compound with a very distinct
behavior.\cite{ref5} In this system, and as a function of y, a
sharp change in the insulator-metal transition temperature T$_{im}$
is obtained in the $y$ region near 0.3. In the high $y$ region, a net
separation between T$_{im}$ and T$_{C}$ can be
observed.\cite{ref15} Below T$_{C}$ the state of the system is
characterized by the coexistence of isolated FM clusters embedded in
charge ordered regions. Unlike what happens in
La$_{0.5}$Ca$_{0.5}$MnO$_{3}$, in  the Pr based compound the FM
fraction increases as the temperature decreases below 80 K, and the
insulator-metal transition is achieved when the fraction of the FM
phase reaches the percolation threshold.\cite{ref15} As $y$
approaches 0.41 from bellow, T$_{im}$ is lower and the maximum of the
MR curve is higher.\cite{ref5} We have studied the MR obtained by
Eq. 2 as a function of the percolation temperature T$_{im}$ by
varying the relation $\rho_{co}/\rho_{FM}$, which is a monotonous function of
T$_{im}$. The analytical description of this situation can be done
within the following scheme: a) At T$_{im}$ the FM fraction is close
to $f_c$=0.17,\cite{ref15} and is fixed by definition for all
samples; b) the temperature T$_{im}$ enters in the calculations
through the ratio $\rho_{co}/\rho_{FM}$, which can change several
orders of magnitude as a function of temperature; c) at T$_{im}$ the
fraction enlargement $\Delta f$ is the same for all samples, and it
is only determined by the value of the magnetic field.  In Fig 3b we
show the results corresponding to the MR values at T$_{im}$ of
different samples obtained from resistivity curves measured with $H$
= 0 and 0.4 T. The raw experimental data was extracted from Refs.
\cite{ref5} and \cite{ref15}, and the MR calculated as a function of
$(\rho_{co}/\rho_{FM})^{1/2}$. The calculations were made taking into
account that the maximum of the MR in Eq. 2 is obtained, for a fixed
$\Delta f$, when $f = f_c - \Delta f/2$. The only free parameter in
the calculations is $\Delta f$ , which was found equal to 0.054; this
corresponds to a value of  $\alpha_f \approx 0.13$ T$^{-1}$, although
the relevance of this coefficient is not proved in
La$_{5/8-y}$Pr$_y$Ca$_{3/8}$MnO$_3$ compounds.  The main fact to be
noted is that only one parameter is sufficient to account for the MR
at T$_{im}$ through several orders of magnitude, the huge changes of
MR being a consequence of the variation of the $\rho_{co}/\rho_{FM}$
ratio as the percolation temperatures change. This can be taken as a
clear indication on behalf of the role played by the magnetic field
in PS systems. \\
\begin{figure}[tbp]
\epsfxsize=8.0cm\centerline{\epsffile{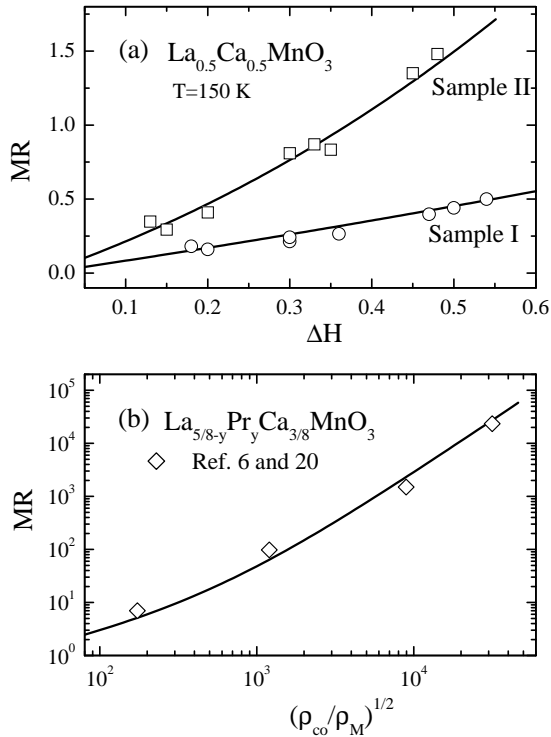}}
\caption{ {\small Fig. 3: Measured (symbols) and calculated (lines) magnetoresistance
MR=$\rho_e(H)/\rho_e(H+\Delta H)$-1 for a) samples I ($f_0$=0.55) and
II ($f_0$=0.09) of La$_{0.5}$Ca$_{0.5}$MnO$_{3}$ at  T = 150 K 
as a function of $\Delta H$; the calculations were done taking $t=2$
and $\rho_{co}/\rho_{FM}$=1000 in Eq. 2; the values $\alpha_f$=0.20
and 0.11 T$^{-1}$ were obtained for samples I and II respectively by
adjusting the experimental data points; 
and b)
La$_{5/8-y}$Pr$_y$Ca$_{3/8}$MnO$_3$ as a function of
$(\rho_{co}/\rho_{FM})^{1/2}$; experimental data points were
extracted from 
Ref. 6 and Ref. 20, corresponding
to samples with  $y$ values of 0.30, 0.35, 0.375 and 0.40; $f_c$ =0.17 was taken in Eq. 2, and $\alpha_f$=0.13 T$^{-1}$ was obtained.}
\label{Fig3}}
\end{figure}

\section{Concluding Remarks} 

We have shown some important aspects
concerning the effects of magnetic fields in PS systems. On one hand,
we have presented evidence that even a low magnetic field has the
capability of modifying the volume fractions of the coexisting phases
in a field-cooling process. This effect seems to be the one that
drives the colossal magnetoresistance observed in these systems. On
the other hand we have shown that the response of the system is
highly dependent on the way in which the magnetic field is applied.
The fact that the large values of MR are obtained only in FC
experiments put strong restrictions to the use of PS systems in
practical applications: magnetic devices taking advantage of the MR
effect are actually exposed to processes like our magnetic field
turn-on-turn-off experiment. In such a case, the MR response of the
PS compound will be rather similar to that obtained in an homogeneous
FM system, i.e., displaying values far away from being considered
"colossal".  Another interesting point related to the measurement
methods we dealt with is the fact that, combining them, we can
determine the onset temperature at which a system gets into the phase
separated state. \\
Finally, using the fraction enlargement picture we
have been able to account for the relation between the FM volume
fraction and the MR in La$_{0.5}$Ca$_{0.5}$MnO$_{3}$, and between MR
and T$_{im}$ in La$_{5/8-y}$Pr$_y$Ca$_{3/8}$MnO$_3$. In this scenario
the fraction expansion coefficient $\alpha_f$ appears as the main
parameter to characterize the PS systems, as it governs the
interrelation between transport properties and magnetic fields. The
interesting question arising from the last point is related to the
dependence of $\alpha_f$ with some physical parameters, namely, $f$, $H$ and
T. We obtained only partial information about that dependence in the
PS systems mentioned here, although the similar values of $\alpha_f$
found in all them, with very different conditions with respect to
$f$, $H$ and T, give some confidence on the physical entity of
$\alpha_f$.  Further experimental and theoretical work is indeed
needed to corroborate this point.\\

We thank Dr. Ruben Weht for his fruitful comments.\\
+  also at EcyT, Universidad Nacional de Gral. San Mart\'{\i}n, San Mart\'{\i}n,
Argentina. \\
++ also at CIC, CONICET, Argentina.\\

\end{document}